\documentstyle[12pt]{article}
\advance \topmargin by -\headheight
\advance \topmargin by -\headsep
     
\evensidemargin \oddsidemargin
\begin{document}
\topmargin -.6in
\def\br{\begin{eqnarray}}
\def\er{\end{eqnarray}}
\def\be{\begin{equation}}
\def\ee{\end{equation}}
\def\nn{\nonumber}
\def\({\left(}
\def\){\right)}
\def\a{\alpha}
\def\b{\beta}
\def\d{\delta}
\def\D{\Delta}
\def\eps{\epsilon}
\def\g{\gamma}
\def\G{\Gamma}
\def\h{ {1\over 2}  }
\def\hp{ {+{1\over 2}}  }
\def\hm{ {-{1\over 2}}  }
\def\k{\kappa}
\def\l{\lambda}
\def\L{\Lambda}
\def\m{\mu}
\def\n{\nu}
\def\o{\over}
\def\O{\Omega}
\def\p{\phi}
\def\rh{\rho}
\def\s{\sigma}
\def\t{\tau}
\def\th{\theta}
\def\ii {\'\i  }
\vskip .6in

\begin{center}
{\large {\bf The Hierarchy of Hamiltonians for a Restricted Class of Natanzon Potentials }} 
\end{center}

\normalsize
\vskip 1cm
\begin{center}

{\it  Elso Drigo Filho $^1$ and Regina Maria  Ricotta $^2$} \\
$^1$ Instituto de Bioci\^encias, Letras e Ci\^encias Exatas, IBILCE-UNESP\\
Rua Cristov\~ao Colombo, 2265 -  15054-000 S\~ao Jos\'e do Rio Preto - SP\\
$^2$  Faculdade de Tecnologia de S\~ao Paulo, FATEC/SP-CEETPS-UNESP  \\
Pra\c ca  Fernando Prestes, 30 -  01124-060 S\~ao Paulo-SP\\ 
Brazil\\
\end{center}
{\bf  Abstract}\\
\baselineskip 0.3in   
\noindent The restricted class of Natanzon potentials with two free parameters is studied within
the context of  Supersymmetric Quantum Mechanics.  The hierarchy of Hamiltonians is indicated,
where the first members of the superfamily are explicitly evaluated and a general  form for the
superpotential is proposed.\\

\noindent {\bf  I. Introduction}\\
\indent The set of potentials known as Natanzon potentials have numerous applications in $\,$
several branches of physics, \cite{Natanzon}-\cite{Cooper2}.  An important point to stress is
that the two classes  of Natanzon potentials, the hypergeometric and the confluent, include all the 
potentials whose Schroendinger equation is analytically and exactly solvable, \cite{Ginocchio1}. 
They have motivated several works concerning the mathematical and algebraic aspects of their
structure and solutions, \cite{Ginocchio1}-\cite{Milson}.

In particular, there are studies within  the context Supersymmetric Quantum Mechanics formalism.
Cooper et al, \cite{Cooper1}, for instance, investigated the relationship between shape
invariance and exactly analytical solvable potentials and showed that the Natanzon potential is
not shape invariant although it has analytical solutions for the associated Schroedinger
equation.  L\'evai et al, \cite{Levai}, have determined phase-equivalent potentials for a class of Natanzon potentials
employing the formalism of supersymmetry. 

However, the hierarchy of the Hamiltonians corresponding to Natanzon potentials has not
been  determined  yet.  Cooper et al in ref. \cite{Cooper2} attempted to this possibility but the
hierarchy was not computed, since they were only interested  in the first two members of the
hierarchy, the two partner Hamiltonians, in order to check the shape invariance.  In this work we
explore this point. The superalgebra is used to construct the hierarchy of Hamiltonians of the
restricted class of Natanzon potentials, with two free parameters.  The first few members of the
superfamily are explicitly evaluated and a general form for the superportential is proposed by induction. 
Comments about exactly solvable potentials and their relationship with supersymmetry are
given in the conclusions.
\\

\noindent {\bf  II. Supersymmetric Quantum Mechanics Formalism}\\
In the formalism of Supersymmetric Quantum Mechanics there are two operators  $Q$ and
$Q^+$, that satisfy the algebra
\be
 \{Q, Q\}  = \{Q^+, Q^+\} = 0,  \;\;\;\;\{ Q, Q^+\} = H_{SS}
\ee
where $H_{SS}$ is the supersymmetric Hamiltonian. The usual realisation of the operators   
$Q$ and $Q^+$ is 
\be
Q =  a_1^- \sigma^- =  \left( \begin{array}{cc} 0  & 0  \\ a_1^-  & 0 
\end{array} \right ) \;,\;\;\;
Q^+ = a_1^+ \sigma^+ = \left( \begin{array}{cc} 0  & a_1^+  \\ 0 & 0 
\end{array} \right )
\ee
where $\sigma^\pm$ are written in terms of the Pauli matrices and  $a_1^{\pm}$ are bosonic
operators written in terms of the superpotential $W_1(r)$:
\be 
\label{a's}
a_1^{\pm} =  \left(\mp {d \o dr} + W_1(r) \right).
\ee
With this realisation the supersymmetric Hamiltonian  $H_{SS}$ is given by
\be
\label{H_SS}
H_{SS} = \left( \begin{array}{cc} a_1^+a_1^-  &  0 \\ 0 & a_1^-a_1^+
\end{array} \right ) = \left( \begin{array}{cc} H_1  &  0 \\ 0 & H_2
\end{array} \right )
\ee
where $H_1$ and $H_2$ are supersymmetric partner Hamiltonians.  There is a direct
relationship between these Hamiltonians and their spectra of eigenfunctions and eigenvalues,  namely,
\br
\Psi_{n+1}^{(1)} \propto a_1^+\Psi_n^{(2)}\;,\;\;\; \Psi_n^{(2)} \propto a_1^-\Psi_{n+1}^{(1)} \nn \\
E_{n+1}^{(1)} = E_n^{(2)}
\er
with $E_0^{(1)} = 0$, ($n = 0, 1, ...$).

The superpotential and the ground state eigenfunction are related by
\be
W_1(r) = -{d\o dr} log(\Psi_0^{(1)}) .
\ee
Substituting $a_1^{\pm}$ as defined in (\ref{a's}) into $H_1$  given by (\ref{H_SS}) we end up with the Riccati
equation satisfied by $W_1$,
\be
W_1^2 - {dW_1\o dr} = V_1(r) - \epsilon_0^{(1)}.
\ee

This structure can be repeated for $H_2$, i.e., $H_2$ can be factorized again in terms of
its  ground state 
\be
H_2 = a_2^+a_2^-
\ee
with
\be
a_2^{\pm} =  \left(\mp {d \o dr} + W_2(r) \right)
\ee
where the superpotential $W_2$ can be written in terms of the ground state eigenfuntion of
$H_2$
\be
W_2(r) = -{d\o dr} log(\Psi_0^{(2)}) .
\ee

Repeating this process $n$ times, we get a whole family of Hamiltonians, related by
supersymmetry, \cite{Cooper2}, \cite{Sukumar}
\be
H_n= a_n^+a_n^-
\ee
with
\be
a_n^{\pm} =  \left(\mp {d \o dr} + W_n(r) \right).
\ee

The supersymmetry enables us to relate all the $n $ members of the hierarchy, as made for
the first two members, see fig.1,
\br
\Psi_n^{(1)} \propto a_1^+a_2^+...a_n^+\Psi_0^{(n+1)} \nn \\
E_n^{(1)} = E_0^{(n+1)}
\er
where the eigunfunctions must be normalizable. \\

\noindent{\bf  III. Natanzon Potential and the Hierarchy of Hamiltonians}\\
The restricted class of Natanzon potentials having two parameters  and  given in terms of the variable $y(r)$ is,
\be
\label{Natanzon} 
V(r) = \{ -\lambda^2 v(v+1) + {1\o 4}(1 - \lambda^2) [5(1 - \lambda^2) y^4 - (7 - 
\lambda^2) y^2 + 2]\} (1 - y^2)\;\;,
\ee     
where the variable function $y(r)$ satisfies 
\be
dy/dr = (1 - y^2)[1 - (1 -
\lambda^2)y^2].
\ee
The dimensionless free parameters $v$ and $\lambda$  measure the depth and the
shape of the potential, respectively. 

We write the Schroedinger equation  for this potential, \cite{Ginocchio1}, \cite{Ginocchio2}, in
dimensionless units $r = bx =(2mv_0/\hbar^2)^{1/2}x $
\be
[-d^2/dr^2 + V(r)]\Psi_n(r) = \epsilon_n \Psi_n(r)
\ee
where $V(x) = v_0 V(r)$, $\epsilon_n = E_n/v_0$ .

The analytic solutions for the energy eigenfunctions  are given by, 
\be
\label{Psin}
\Psi_n = (1 - \lambda^2)^{\mu_n/2} [g(y)]^{-(2\mu_n +1)/4} C^{\mu_n +1/2}_n(\lambda
y/[g(y)]^{1/2})
\ee
where $g(y) = 1 - (1 - \lambda^2)y^2$.   The factor $C^{(a)}_n(x)$ is a Gegenbauer polynomial when $n$ is a
non-negative integer, which is our case. The correponding energy eigenvalues are given by 
\be
\epsilon_n = - \mu_n^2 \lambda^4\;, \;\;\;\mu_n > 0, 
\ee
where
\be
\mu_n\lambda^2 = [\lambda^2 (v+1/2)^2 + (1 - \lambda^2)(n+1/2)^2]^{1/2} - (n + 1/2). 
\ee
Notice the relationship between the energy levels which will be extensively used in what follows,
\be
(\mu_n^2 - \mu_{n-1}^2)\lambda^2 =  ( -2n  - (2n + 1)\mu_n + (2n - 1)\mu_{n-1}).
\ee
In order to construct the superfamily we firstly factorize the Natanzon potential,
calling $V(r) = V_1(r) = V_-(r) + \epsilon_0^{(1)}$, \cite{Cooper1}, whose Schroedinger equation is 
\be 
H_1 - \epsilon_0^{(1)} = a_1^+ a_1^- \;,\;\;\; a_1^{\pm} = \mp {d \o dr} + W_1(r)
\ee
where $\epsilon_n^{(1)} = \epsilon_n$.  The superpotential $W_1(r)$ is evaluated from the knowledge of the
ground-state eigenfunction of $V(r)$, $W_1(r) = -{d\o dr} log(\Psi_0^{(1)}) $, where $\Psi_n^{(1)} = \Psi_n $,
given  by (\ref{Psin}). It satisfies the Riccati equation and it is given by
\be
W_1(r) = {1\o 2} (1- \lambda^2) y (y^2 - 1) + y \mu_0 \lambda^2.
\ee
The superpartner Hamiltonian satisfies the equation 
\be 
H_2 - \epsilon_0^{(1)} = a_1^- a_1^+ 
\ee
which is written in terms of $V_2(r)$ like
\be
W_1^2 + {dW_1\o dr} = V_2(r) - \epsilon_0^{(1)}
\ee
where 
\be
\label{V2} 
V_2(r) = \{ -\mu_0^2 \lambda^4 + \mu_0 \lambda^2 + {1\o 4}(1 - \lambda^2) [- 7(1 -
\lambda^2) y^4 + (9 - 3\lambda^2 - 8\mu_0 \lambda^2) y^2 - 2]\} (1 - y^2).
\ee     
To construct the next member of the superfamily, we factorize the Schroedinger equation for $V_2$. It gives
\be 
H_2 - \epsilon_0^{(2)} = a_2^+ a_2^- \;,\;\;\; a_2^{\pm} = \mp {d \o dr} + W_2(r)
\ee
where $W_2(r)$ satisfies the associated Riccati equation,
\be
W_2^2 - {dW_2\o dr} = V_2(r) - \epsilon_0^{(2)}\;.
\ee
$\epsilon_0^{(2)}$ is the energy  ground state of the potential $V_2(r)$ and it
is such that   $\epsilon_0^{(2)} = \epsilon_1^{(1)}$, see fig. 1. The superpotential $W_2$ is given
by $W_2(r) = -{d\o dr} log(\Psi_0^{(2)})$, where $\Psi_0^{(2)} = a_1^-\psi_1^{(1)}$, i.e.,
\be
W_2(r) = {3\o 2} (1- \lambda^2) y (y^2 - 1) + y \mu_1 \lambda^2 -{d\o dr} log(f_1)
\ee
where 
\be
\label{f1}
f_1(y) = 1 + a_{11} y^2\;, \;\;\; a_{11} = (\mu_0 - \mu_1)\lambda^2 - 1.
\ee
The new superpartner of $H_2$ is given by
\be
W_2^2 + {dW_2\o dr} = V_3(r) - \epsilon_0^{(2)}
\ee
where 
\br
\label{V3} 
V_3(r)&=& \{ -\mu_1^2 \lambda^4 + \mu_1 \lambda^2 + {1\o 4}(1 - \lambda^2) \left(-27(1 -
\lambda^2) y^4 + (33 - 15 \lambda^2) y^2 - 6\right) + \\
& +&  {2 a_{11} g(y) \o f_1(y)} \{1 + (- 9 + 6 \l^2 - 2 \mu_1  \l^2) y^2 + 8 (1 - \l^2)
y^4\} + 8a_{11}^2 y^2 (1 - y^2)({g(y)\o f_1(y)})^2 \} (1 - y^2).\nn 
\er     
and $g(y) = 1 -  (1- \lambda^2) y^2$.
Thus, factorizing the Hamiltonian for this potential we have
\be 
H_3 - \epsilon_0^{(3)} = a_3^+ a_3^- \;\;\;\;, a_3^{\pm} = \mp {d \o dr} + W_3(r)
\ee
where $W_3(r)$ satisfies the Riccati equation,
\be
W_3^2 - {dW_3\o dr} = V_3(r) - \epsilon_0^{(3)}\;.
\ee
$\epsilon_0^{(3)}$ is the energy  ground state of the potential $V_3(r)$, with   $\epsilon_0^{(3)} 
 = \epsilon_1^{(2)} =\epsilon_2^{(1)}$, see fig. 1.  The superpotential, defined by $W_3(r) = -{d\o dr}
log(\Psi_0^{(3)})$ and
$\Psi_0^{(3)} = a_2^-a_1^-\psi_2^{(1)}$, is given by
\be
W_3 = {5\o 2} (1- \lambda^2) y (y^2 - 1) + y \mu_2 \lambda^2 + {d\o dr} log(f_1) - {d\o
dr} log(f_2)
\ee
where 
\be
\label{f2}
f_2(y) = 1 + a_{21} y^2 + a_{22} y^4
\ee
 with
\be
a_{21} = 2(\mu_1 - \mu_2)\lambda^2 - 2   
\ee
\br
a_{22} = 1 + {\lambda^2 \o 3}(2 - \mu_0 - 3 \mu_1 + 6 \mu_2) + {\lambda^4 \o 3}(-4 \mu_0 + 6\mu_1 - 2\mu_2 -
\mu_0^2  - 5
\mu_0 \mu_2 + 3 \mu_0 \mu_1 + 3\mu_1 \mu_2) . \nonumber
\er
For the next member of the superfamily, we show the result of the evaluation of the superpotential $W_4(r) = -{d\o
dr} log(\Psi_0^{(4)})$ with
$\Psi_0^{(4)} = a_3^- a_2^- a_1^-\psi_3^{(1)}$. It is given  by
\be
W_4 = {7\o 2} (1- \lambda^2) y (y^2 - 1) + y \mu_3 \lambda^2 + {d\o dr} log(f_1) + {d\o
dr} log(f_2) - {d\o dr} log(f_3)
\ee
where 
\be
\label{f3}
f_3(y) = 1 + a_{31} y^2 + a_{32} y^4 + a_{33} y^6
\ee
 with
\be
a_{31} =3(\mu_2 - \mu_3)\lambda^2 +  2(\mu_0 - \mu_1)\lambda^2 -5  \nonumber
\ee
\br
a_{32} =  10  &+  & {\lambda^2 \o 3}(6  + 25 \mu_0 + 15 \mu_1 - 6 \mu_2 + 22 \mu_3) +  \nonumber \\ 
& + &{\lambda^4 \o 3}( 2 \mu_0  + 2\mu_0^2 - 18 \mu_1  - 6 \mu_0 \mu_1 + 18 \mu_2  + 13 \mu_0 \mu_2  -
\\
 & - &  3 \mu_1 \mu_2 - 6 \mu_3 - 11\mu_0 \mu_3 - 3 \mu_1 \mu_3 + 8 \mu_2 \mu_3 )   \nonumber 
\er
\br
a_{33} =  -10 &+ & {\l^2} (-6+13  \mu_0-3  \mu_1-12  \mu_2-4  \mu_3)+{\l^4} \Big(-{148\o 15}+{484  \o
45}\mu_0-  {206  \o 45} \mu_0^2-{84  \o 5}\mu_1+\nonumber  \\
&+&{42 \o 15}  \mu_0
\mu_1 + 2  \mu_2  +{1\o 3}\mu_0  \mu_2- \mu_1  \mu_2- 
{394  \o 45}\mu_3 +{497 \o 45} \mu_0 \mu_3- {59  \o 5}\mu_1  \mu_3- {16  \o 3}\mu_2  \mu_3 \Big) +
 \nonumber \\  
& +& {\l^6} \Big({8 \o 3} \mu_0+{328 \o 45} \mu_0^2-{116 \o 45} \mu_0^3-{24 \o 5}
\mu_1- {48  \o 5 }\mu_0  \mu_1+ {12 \o 5} \mu_0^2  \mu_1+ {8  \o 3}\mu_2  + {82  \o
9}\mu_0  \mu_2 - \nonumber \\  
&-&{14 \o 9} \mu_0^2 \mu_2 -6  \mu_1  \mu_2 +2  \mu_0  \mu_1  \mu_2- {8  \o 15}\mu_3 - {58 \o 45} \mu_0 
\mu_3  +  {182 \o 45} \mu_0^2  \mu_3-{2 
\o 5}\mu_1  \mu_3 - \nonumber \\ 
& -&  {26 \o 5} \mu_0  \mu_1  \mu_3 + {8  \o 9}\mu_2 \mu_3 + {44 \o 9} \mu_0 
\mu_2  \mu_3 -4 
\mu_0 \mu_2 \mu_3\Big) 
\er

Casting all the results we have so far for the hierarchy, the following nth-term for the superpotential is induced
\be
W_{n+1} = {2n + 1 \o 2}(1- \lambda^2) y (y^2 - 1) + y \mu_n \lambda^2 + {d\o dr}
log({\prod_{i=0}^n f_{i-1} \o f_n})\;\;\;,\; f_0 = f_{-1} = 1
\ee
where $f_n(y)$ is a $2n$-order polynomial  of the form
\be
f_n(y) = \sum_{i=0}^n a_{in} y^{2i} \;\;\;,\; a_{n0} = 1.
\ee

We stress that since $W_{n+1}$  is a superpotential it checks the Riccati equation,
\be
W_{n+1}^2 - {dW_{n+1}\o dr} = V_{n+1}(r) - \epsilon_0^{(n+1)} \nonumber
\ee
where $V_{n+1}(r)$ is the superpartner potential of $V_n$ which satisfies 
\be
W_n^2 + {dW_n\o dr} = V_{n+1}(r) - \epsilon_0^{n}. \nonumber
\ee
We have therefore a recursive relationship between $W_{n+1}$ and $W_{n}$ given by
\be
W_{n+1}^2 - {dW_{n+1}\o dr} = W_n^2 + {dW_n\o dr} + \epsilon_0^{n} - \epsilon_0^{(n+1)}
\ee
where $\epsilon_0^{n} = - \mu_{n -1}^2 \lambda^4$  and  $ \epsilon_0^{(n+1)}= - \mu_{n }^2 \lambda^4$.
After the  substitutions  we end up with the condition
\br
\lefteqn{2n (1 - \l^2)^2 y^2 (y^2  -1)^2 + (y^2  -1) \l^2 \left(((2n - 1) \mu_{n -1} - (2n  + 1) \mu_n) (1 - (1 - \l^2)
y^2) - 2n \right) +} \nonumber \\
& &+ \sum_{i=0}^{n-1}{f_{i-1}^{\prime}\o f_{i-1}} \left( 4 {f_{n-1}^{\prime}\o f_{n-1}} - 2 {f^{\prime}_n\o f_n} + 2
(1 - \l^2) y (y^2  -1) + 2 y \l^2 (\mu_n - \mu_{n - 1}) \right) +  \nonumber \\
& &+ {f^{\prime}_{n -1}\o f_{n -1}} \left(  4 n  (1 - \l^2) y (y^2  -1) - 2 {f^{\prime}_n\o f_n} + 2 y \l^2 (\mu_n +
\mu_{n - 1})\right) -  \\
& & - {f^{\prime}_n\o f_n}\left(  (1 - \l^2) y (y^2  -1)(2n + 1) + 2 y \l^2 \mu_n
\right) + \nonumber \\
& &+ {f^{\prime \prime}_n\o f_n} - 2 \sum_{i=0}^{n-1}{f_{i-1}^{\prime \prime}\o f_{i-1}} + 2
\sum_{i=0}^{n-1}({f_{i-1}^{\prime}\o f_{i-1}})^2 + \left( 2n (1 - \l^2) (1 - 3 y^2)  - \l^2 (\mu_n + \mu_{n - 1})
\right){d y \o dr }
 = 0 .\nonumber
 \er
where $f^\prime = {d f \o dr }$ and $f^{\prime \prime} = {d^2 f \o dr^2 }$.

Therefore,  $f_{n+1}$ can be determined from the knowledge of  $f_n $. In  this way, the particular cases of  $n=1$,
$n=2$ and $n=3$ can be checked by inspection and the resulting functions
$f_1$, $f_2$ and $f_3$ perfectly agree with equations (\ref{f1}) , (\ref{f2}) and (\ref{f3}) respectively.
\\
\begin{figure}
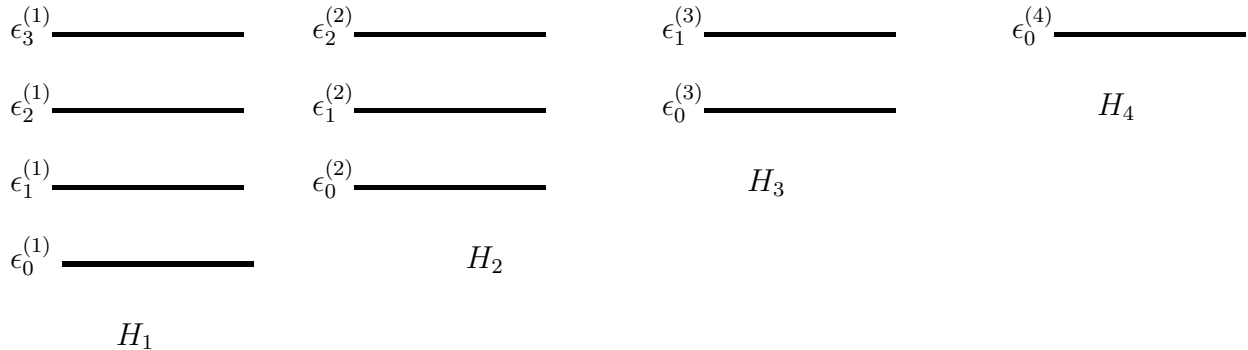

$\epsilon_3^{(1)}$\noindent \rule[.01in]{1in}{.02in}  \hspace{.25in} 
$\epsilon_2^{(2)}$\rule[.01in]{1in}{.02in}
\hspace{.5in}  $\epsilon_1^{(3)}$\rule[.01in]{1in}{.02in} \hspace{.5in} 
$\epsilon_0^{(4)}$\rule[.01in]{1in}{.02in} \\

$\epsilon_2^{(1)}$\noindent \rule[.01in]{1in}{.02in} \hspace{.25in} 
$\epsilon_1^{(2)}$\rule[.01in]{1in}{.02in} \hspace{.5in} 
$\epsilon_0^{(3)}$\rule[.01in]{1in}{.02in}\hspace{1in} $H_4$\\

$\epsilon_1^{(1)}$\noindent \rule[.01in]{1in}{.02in} \hspace{.25in}
$\epsilon_0^{(2)}$\rule[.01in]{1in}{.02in}\hspace{1in} $H_3$\\

$\epsilon_0^{(1)}$ \noindent \rule[.01in]{1in}{.02in} \hspace{1in} $H_2$\\

\hspace{.5in} $H_1$
\caption{The hierarchy scheme}
\end{figure}

\noindent {\bf IV. Conclusions}\\
The hierarchy of Hamiltonians is studied for the restricted class of Natanzon potentials, with
two parameters and a general form for the superpotential is proposed.  The
superalgebra drives us to the conclusion that the whole superfamily is a collection of exactly
solvable Hamiltonians.

As a final remark, some aspects related to  shape  invariance emerge from the results
presented here.  The shape invariance concept introduced by Gedenshtein, \cite{Gedenshtein},
has motivated several discussions about the exactly solvable potentials. In ref. \cite{Cooper2}
there is an extensive explanation about this subject.

The Natanzon potential is not shape invariant in the Gedenshtein sense, \cite{Cooper1}. 
However, for the restricted class analised here, it was possible to obtain a general form for the
superpotential, as shown in the previous section.  Thus,  we conceived a more general
condition, written in terms of the whole superfamily (not in terms of two members, $V_1$
and $V_2$, as usual), to indicate if a given potential is exactly solvable or not.

We argue that a criteria of solvability of a potential can be written in terms of the
superpotential.  In this way, if it is possible to construct a general expression for all
superpotentials in the hierarchy of Hamiltonians then the original potential is exactly solvable.

The Hulth\'en potential without the potential barrier term is another example of an \-exactly solvable potential which  is
not shape invariant, but  for which  it is possible to determine a general expression for the superpotential in the
hierarchy,  \cite{Drigo2}.\\

\noindent {\bf  Acknowledgments}\\
E. D. F. would like to thank CNPq agency (Brazil) for partial financial support.\\

\end{document}